\documentclass[10pt,amsmath,amssymb,graphicx,nofootinbib]{revtex4} 
\usepackage[english]{babel}
\usepackage[latin1]{inputenc}
\usepackage{amsmath}
\usepackage{bbm}
\usepackage{graphicx}
\usepackage{bm}
\usepackage{graphics,graphicx,calc,epsfig,pstricks,bbm}
\usepackage{amsmath,amssymb,amsfonts}
\newcommand{\be}{\begin{equation}}
\newcommand{\ee}{\end{equation}}
\newcommand{\bea}{\begin{eqnarray}}
\newcommand{\eea}{\end{eqnarray}}

\begin{document}
\title{Diffusion on $\kappa$-Minkowski space}

\author{Michele Arzano}
\email{michele.arzano@roma1.infn.it}
\affiliation{Dipartimento di Fisica and INFN,\\ ``Sapienza" University of Rome,\\ P.le A. Moro 2, 00185 Roma, Italy}

\author{Tomasz Trze\'{s}niewski}
\email{tomasz.trzesniewski@ift.uni.wroc.pl}
\affiliation{Institute for Theoretical Physics,\\ University of Wroc\l{}aw,\\ Pl.\ Maxa Borna 9, 50-204 Wroc\l{}aw, Poland}

\begin{abstract}
\begin{center}
{\bf Abstract}\\
\end{center}
We study the spectral dimension associated with diffusion processes on Euclidean $\kappa$-Minkowski space. We start by describing a geometric construction of the ``Euclidean" momentum group manifold related to $\kappa$-Minkowski space. On such space we identify various candidate Laplacian functions, i.e.\! deformed Casimir invariants, and calculate the corresponding spectral dimension for each case. The results obtained show a variety of running behaviours for the spectral dimension according to the choice of deformed Laplacian, from dimensional reduction to super-diffusion.
\end{abstract}


\maketitle
\section{Introduction}
The notion of {\it spectral dimension} has played an important role in recent years as a way of characterizing (quantum) geometries beyond the usual picture of smooth manifolds, emerging in various quantum gravity scenarios \cite{Ambjorn:2005db,Lauscher:2005qz,Horava:2009if,Modesto:2008jz,Carlip:2009kf,
Benedetti:2009ge,Sotiriou:2011aa,Calcagni:2013vsa}. A typical feature, common to almost all models, is that of a spectral dimension running from a value coinciding with the Hausdorff dimension of 4 in the infrared to the value of 2 in ultraviolet regimes. Motivated by this quantum gravitational {\it dimensional reduction} various models beyond local quantum field theory have been proposed and studied which exhibit a scale-dependent spectral dimension \cite{Calcagni:2010bj,Arzano:2011yt,Alesci:2012,Amelino:2013}. Non-commutative field theories with momentum space given by a non-abelian group manifold are among such models. These are characterized by non-commuting space-time coordinates closing the Lie algebra generating the momentum Lie group and by a ``quantum deformation" of ordinary relativistic symmetries, which captures the non-trivial structure of momentum space.\\
One example of such theories is encountered in the study of point particles coupled to Einstein gravity in three space-time dimensions. Such particles are described by topological defects whose momenta live on the Lorentz group. As shown in \cite{Alesci:2012} such Lorentz group momentum space leads to non-trivial diffusion. In particular the behaviour of the diffusion process is determined by the spherical geometry of the Euclidean momentum space and the spectral dimension approaching UV scales is initially super-diffusive and at sub-Planckian spaces drops to zero.\\
An interesting question is whether analogous models in four space-time dimensions lead to non-trivial diffusion behaviours. The only known example of deformed symmetries linked to a Lie group momentum space in four dimensions is given by the $\kappa$-Poincar\'e algebra and related $\kappa$-Minkowski non-commutative space-time \cite{Lukierski:1991,Lukierski:1994,Majid:1994}. Here the momentum space is described by a Lie group whose manifold covers half de Sitter space \cite{Kowalski:2002,Kowalski:2003,Nowak:2002}. The behaviour of the spectral dimension for $\kappa$-Minkowski space was first studied in \cite{Benedetti:2009} where it was shown that under certain prescription for defining an Euclidean version of the $\kappa$-Poincar\'e algebra and under a specific choice of the deformed Laplacian operator a running of the spectral dimension from 4 to 3 in the UV is observed. A more geometric approach was adopted in the recent \cite{Amelino:2013} which studied the {\it asymptotic behaviour} of the spectral dimension associated with a de Sitter momentum space for several possible choices of the Laplacian. It was showed that indeed the running of the spectral dimension to a value of 2 in the UV can be realized by a particular choice of the Laplacian.\\
In this work we present a systematic study of diffusion and running spectral dimension in Euclidean $\kappa$-Minkowski space. We start in the next Section by proposing a novel ``geometric" construction of Euclidean $\kappa$-Poincar\'e momentum group manifold. Our prescription offers a general definition of Euclidean $\kappa$-momentum space and $\kappa$-Poincar\'e translation generators which does not rely on a specific choice of coordinates on the group manifold or basis of generators. In the following Section III we discuss the Hopf algebra structure of the Lorentzian and Euclidean $\kappa$-Poincar\'e algebra with emphasis on the classical and bicrossproduct bases. In Section IV we define diffusion on the Euclidean $\kappa$-Minkowski and calculate the spectral dimension for various choices of the Laplacian operator. In particular we calculate the spectral dimension {\it analytically} for the Laplacian dictated by bi-covariant differential calculus on $\kappa$-Minkowski in three and four space-time dimensions. In the latter case we recover the results obtained numerically in \cite{Benedetti:2009}. We then focus on the popular {\it bicrossproduct} Laplacian associated with the mass Casimir of the $\kappa$-Poincar\'e algebra in the bicrossproduct basis and again calculate the spectral dimension analytically in four and numerically in three dimensions. In the former case we recover the asymptotic values found in \cite{Amelino:2013}. Finally, in the spirit of the ``relative locality" interpretation \cite{Amelino:2011,Kowalski:2013} (see also \cite{Freidel:2014,Leigh:2014}, which provide a link between this framework and the string theory) of curved $\kappa$-Poincar\'e momentum space we consider the Laplacian obtained from the geodesic distance from the origin on the momentum group manifold. In this case we resorted to numerical techniques and found a {\it super-diffusive} behaviour leading to a divergence of the spectral dimension for diffusion scales approaching zero. In Section V we present a summary of our findings and engage in a comparative discussion with similar results from other models.

\section{$\kappa$-Minkowski space and Euclidean $AN(n)$ momentum group manifold}
The $n+1$-dimensional Lie algebra with brackets 
\be\label{kMink}
[X_0, X_a] = \frac{i}{\kappa} X_a\,, \quad [X_a, X_b] = 0\,; \quad a = 1,\ldots,n
\ee
is known in the literature on non-commutative field theory as the $\kappa$-Minkowski {\it non-commutative space-time}. The constant $\kappa$ carries dimension of energy so that, in natural units, the generators have dimension of length. The parameter $\kappa$ is usually taken as a UV scale and often identified with the Planck energy so that at small energy scales, $\kappa\rightarrow +\infty$, commuting coordinates of ordinary  $n+1$-dimensional Minkowski space are recovered. The algebra (\ref{kMink}) can be represented in terms of $(n+2) \times (n+2)$ matrices
\begin{align}\label{eq:2.1}
X_0 = -\frac{i}{\kappa} \left(
\begin{array}{ccc}
0\! & {\mathbf 0}\! & 1 \\ 
{\mathbf 0}\! & {\mathbf 0}\! & {\mathbf 0} \\ 
1\! & {\mathbf 0}\! & 0 
\end{array}
\right)\,, \qquad 
X_a = \frac{i}{\kappa} \left(
\begin{array}{ccc}
0\! & {\mathbf e}_a^\top\! & 0 \\ 
{\mathbf e}_a\! & {\mathbf 0}\! & {\mathbf e}_a \\ 
0\! & -{\mathbf e}_a^\top\! & 0 
\end{array}
\right)\,,
\end{align}
where the $n$-vector ${\mathbf e}_a = (0,\ldots,1,\ldots,0)$ has entry $1$ at the $a$'th position. From the matrix representation it is easy to show that the {\it spatial} generators are nilpotent i.e.\! in this case $X_a^3 = 0$ (independently of $n$) and thus the algebra is called the $n$-dimensional {\it abelian nilpotent} algebra and denoted by $\mathfrak{an}(n)$.\\
The Lie group $AN(n)$ obtained by exponentiating the $\kappa$-Minkowski algebra is naturally associated with a momentum space if one interprets exponentials of the coordinates as non-commutative plane waves. A choice of {\it normal ordering} for plane waves corresponds to a parametrization of a group element. In particular we choose to parametrize an element of $AN(n)$ as the ``right-ordered" exponential
\be\label{Kexp}
g = e^{-i k^a X_a} e^{i k_0 X_0}\,.
\ee 
In this convention the real parameters $k_0$, $k_a$ are called the bicrossproduct coordinates \cite{Majid:1994}. In matrix representation the generic group element reads
\begin{align}\label{eq:2.2}
g = \left(
\renewcommand{\arraystretch}{1.5}\begin{array}{ccc}
\cosh\left(\frac{k_0}{\kappa}\right) + e^{k_0/\kappa} \frac{k_ak^a}{2\kappa^2}\! & \tfrac{1}{\kappa} \mathbf{k}^\top\! & \sinh\left(\frac{k_0}{\kappa}\right) + e^{k_0/\kappa} \frac{k_ak^a}{2\kappa^2} \\ 
e^{k_0/\kappa} \tfrac{1}{\kappa} \mathbf{k}\! & \mathbbm{1}\! & e^{k_0/\kappa} \tfrac{1}{\kappa} \mathbf{k} \\ 
\sinh\left(\frac{k_0}{\kappa}\right) - e^{k_0/\kappa} \frac{k_ak^a}{2\kappa^2}\! & -\tfrac{1}{\kappa} \mathbf{k}^\top\! & \cosh\left(\frac{k_0}{\kappa}\right) - e^{k_0/\kappa} \frac{k_ak^a}{2\kappa^2} 
\end{array}
\right)\,,
\end{align}
where $\mathbbm{1}$ denotes $n \times n$ identity matrix and $\mathbf{k}$ the $n$-vector $(k_1,\ldots,k_n)$. An important point to stress is that $\mathfrak{an}(n)$ is actually a subalgebra of the Lorentz algebra $\mathfrak{so}(n+1,1)$ and hence $AN(n)$ is a subgroup of the Lorentz group $SO(n+1,1)$. In particular we have $X_0 = \frac{1}{\kappa} J_{0,n+1}$, $X_a = \frac{1}{\kappa} (J_{a0} + J_{a,n+1})$, where $J_{\mu\nu}$ are rotations and boosts generators i.e.\! elements of the Lorentz algebra $\mathfrak{so}(n+1,1)$. Thus $g$ is an element of the isometry group of the $(n+1)+1$-dimensional Minkowski spacetime.\\
In order to describe the manifold structure of $AN(n)$ let us recall how its elements can be mapped to points in a $n+1$-dimensional de Sitter space \cite{Kowalski:2002}, \cite{Kowalski:2003}. The mapping can be obtained by acting with the group element (\ref{eq:2.2}) on the spacelike vector $(0,\ldots,0,\kappa)$ of the $(n+1)+1$-dimensional Minkowski space. Writing the resulting group element in terms of ``cartesian", or embedding coordinates $g \triangleright (0,\ldots,0,\kappa) = (p_0,\{p_a\},p_{-1})$ we obtain
\begin{align}\label{eq:2.3}
p_0 & = \kappa \sinh\left(\tfrac{k_0}{\kappa}\right) + \frac{1}{2\kappa} e^{k_0/\kappa} k_ak^a\,, \nonumber\\ 
p_a & = e^{k_0/\kappa} k_a\,, \nonumber\\ 
p_{-1} & = \kappa \cosh\left(\tfrac{k_0}{\kappa}\right) - \frac{1}{2\kappa} e^{k_0/\kappa} k_ak^a\,.
\end{align}\\
It is easily verified that the coordinates given above satisfy the following constraints $-p_0^2 + p_ap^a + p_{-1}^2 = \kappa^2$ and $p_0 + p_{-1} > 0$. The first of them defines a $(n,1)$-hyperboloid i.e.\! de Sitter space embedded in Minkowski space. The other condition restricts us to half of the manifold. In order to single out a $n+1$-dimensional energy and momentum element out of the $(n+1)+1$ coordinates $(p_0,\{p_a\},p_{-1})$ let us notice that taking the ``classical limit" $\kappa \rightarrow +\infty$ we obtain $p_0 \rightarrow k_0$, $p_a \rightarrow k_a$ but $p_{-1} \rightarrow +\infty$ and therefore $p_{-1}$ can be taken as the auxiliary coordinate in embedding space depending on other coordinates via the hyperboloid condition. This is the usual ``Lorentzian realization" of the group manifold $AN(n)$ given by half of de Sitter space (cf.\! Fig.\! 1). Notice, however, that one can also map the group to the other half of de Sitter space by replacing the action of $g$ with
\begin{align}\label{eq:2.4}
g {\cal N} \equiv g\cdot\left(
\begin{array}{ccc}
-1\! & {\mathbf 0}\! & 0 \\ 
{\mathbf 0}\! & {\mathbbm 1}\! & {\mathbf 0} \\ 
0\! & {\mathbf 0}\! & -1 
\end{array}
\right)\,.
\end{align}
This just reflects the fact that the $AN(n)$ group can be obtained from the Iwasawa decomposition \cite{KowalskiGlikman:2004tz} of the Lorentz group: $SO(n+1,1) = AN(n) SO(n,1) \cup AN(n) {\cal N} SO(n,1)$. 
\begin{figure}[ht]\label{fig:kdS}
\centering
\includegraphics[width=0.5\textwidth]{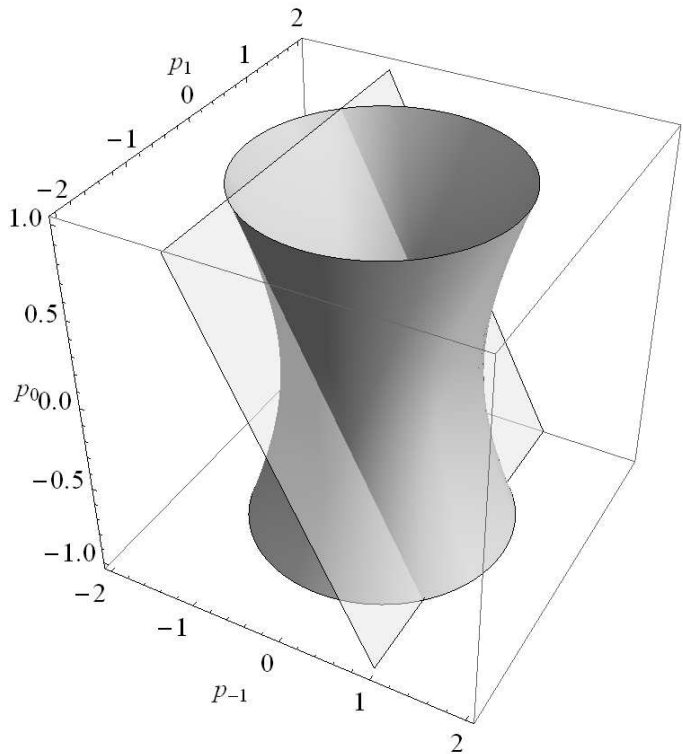}
\caption{Lorentzian de Sitter space of momenta ($p_2,\ldots,p_n$ suppressed) and the $p_0 + p_{-1} = 0$ surface}
\end{figure}
Indeed the full $n+1$-dimensional de Sitter space is equivalent to the quotient $SO(n+1,1) / SO(n,1)$.\\
Since our purpose is to study diffusion processes based on momentum space Laplacians constructed out of coordinates of the $AN(n)$ manifold, we need a prescription to obtain an Euclidean version of the group manifold. We follow here a suggestion from \cite{Kowalski:2013} and map the group $AN(n)$ to an Euclidean manifold by acting with a group element on the {\it timelike} vector $(\kappa,0,\ldots,0)$. We express again the resulting group element as $g \triangleright (\kappa,0,\ldots,0) = (p_{-1},\{p_a\},p_0)$ where now
\begin{align}\label{eq:2.5}
p_0 & = \kappa \sinh\left(\tfrac{k_0}{\kappa}\right) - \frac{1}{2\kappa} e^{k_0/\kappa} k_ak^a\,, \nonumber\\ 
p_a & = e^{k_0/\kappa} k_a\,, \nonumber\\ 
p_{-1} & = \kappa \cosh\left(\tfrac{k_0}{\kappa}\right) + \frac{1}{2\kappa} e^{k_0/\kappa} k_ak^a\,.
\end{align}
As it is easily checked the coordinates now satisfy the conditions $p_0^2 + p_ap^a - p_{-1}^2 = -\kappa^2$ and $p_{-1} > 0$ as well as $p_0 + p_{-1} > 0$, $-p_0 + p_{-1} > 0$ but the latter ones are actually redundant. The first condition defines a $(1,n)$-hyperboloid i.e.\! {\it Euclidean anti-de Sitter space}, also known as hyperbolic space, embedded in Minkowski space. The second condition is again restricting us to one half of the manifold. The subtle point here is that the roles of the coordinates $p_0$ and $p_{-1}$ are now reversed compared to the previous, Lorentzian, case. Indeed in the classical limit $\kappa \rightarrow +\infty$ we obtain $p_0 \rightarrow k_0$, $p_a \rightarrow k_a$ but $p_{-1} \rightarrow +\infty$, which justifies our designation of the coordinates. This is our ``Euclidean realization" of the group $AN(n)$, given as a manifold by half of Euclidean anti-de Sitter space (cf.\! Fig.\! 2). The other half manifold can be obtained in the same way as in the de Sitter case. This reflects another Iwasawa decomposition of the Lorentz group $SO(n+1,1) = AN(n) SO(n+1) \cup AN(n) {\cal N} SO(n+1)$. The full $n+1$-dimensional Euclidean anti-de Sitter space is equivalent to the quotient $SO(n+1,1) / SO(n+1)$. To make an analogy with more familiar structures one can notice that the manifolds we described can be seen as higher dimensional analogues of the four-dimensional mass-shells of tachyons, for the Lorentzian case and of massive particles, for this latter Euclidean realization.
\begin{figure}[ht]\label{fig:kAdS}
\centering
\includegraphics[width=0.5\textwidth]{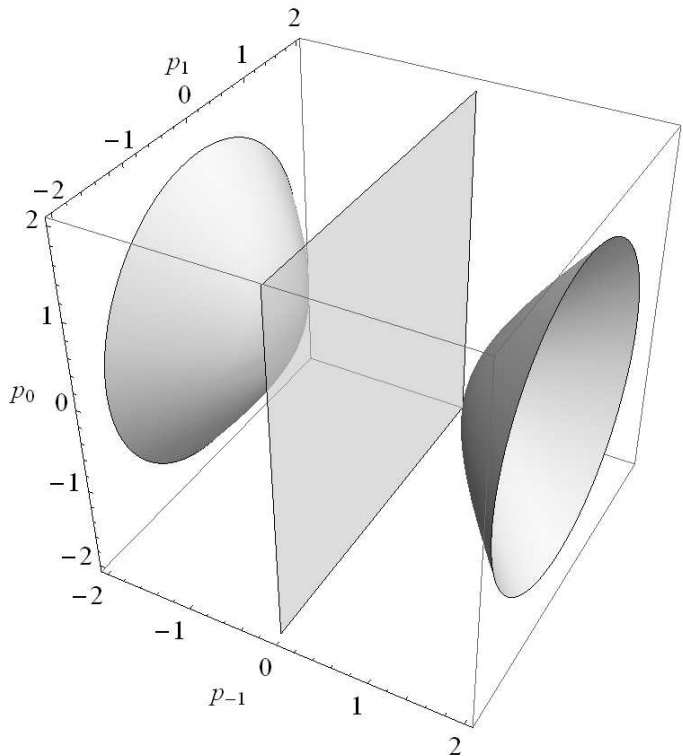}
\caption{Euclidean anti-de Sitter space of momenta ($p_2,\ldots,p_n$ suppressed) and the $p_{-1} = 0$ surface}
\end{figure}\\
A direct connection with the Lorentzian realization can be made by showing that the Euclidean manifold can be obtained by a ``Wick rotation" of the Lorentzian manifold. One can observe this in two steps. First, taking $\kappa \mapsto i \kappa$, $k_0 \mapsto i k_0$ we obtain
\begin{align}\label{eq:2.6}
p_0 & = i \left(\kappa \sinh\left(\tfrac{k_0}{\kappa}\right) - \frac{1}{2\kappa} e^{k_0/\kappa} k_ak^a\right)\,, \nonumber\\ 
p_a & = e^{k_0/\kappa} k_a\,, \nonumber\\ 
p_{-1} & = i \left(\kappa \cosh\left(\tfrac{k_0}{\kappa}\right) + \frac{1}{2\kappa} e^{k_0/\kappa} k_ak^a\right)\,,
\end{align}
satisfying $-p_0^2 + p_ap^a + p_{-1}^2 = -\kappa^2$. Then, taking $p_0 \mapsto i p_0$, $p_{-1} \mapsto i p_{-1}$ we arrive at the ``Euclidean realization".\\
The construction we presented provides a new prescription for obtaining an Euclidean version of the momentum space associated with $\kappa$-Minkowski. Other possible definitions have appeared in the literature. In particular a Wick rotation to imaginary energy and $\kappa$ parameter, $\kappa \mapsto i \kappa$, $k_0 \mapsto i k_0$, was proposed in earlier works on $\kappa$-Poincar\'e \cite{Lukierski:1994} and used to study the spectral dimension \cite{Benedetti:2009}. We see that our prescription on bicrossproduct coordinates coincides with the latter and indeed as we will see our results for the spectral dimension reproduce the numerical result of \cite{Benedetti:2009}. The ``geometric" derivation of Euclidean momentum space we have proposed has the advantage of not relying on a particular set of coordinates on the group manifold. Before moving to the study of the spectral dimension we will see below how the Euclidean $AN(n)$ momentum space is connected to the translation sector of the Euclidean $\kappa$-Poincar\'e algebra.

\section{The $\kappa$-Poincar\'e algebra}
In a scenario in which ordinary momentum $n$-vectors are replaced by elements of a group manifold it is natural to expect that the structure of translation generators and their relation to Lorentz transformations will undergo some non-trivial changes. This is indeed the case for the $AN(n)$ momentum group manifold and the new features which emerge will lead us to describe a quantum deformation of the Poincar\'e algebra known as the $\kappa$-Poincar\'e algebra.\\
The most obvious change compared to ordinary relativistic structures reflects the non-abelian nature of momentum space. Indeed we can express the product of two group elements $g = e^{-i k^a X_a} e^{i k_0 X_0}$ and $h = e^{-i l^a X_a} e^{i l_0 X_0}$ as 
\be\label{eq:3.1a}
g h = e^{-i (k^a \oplus l^a) X_a} e^{i (k_0 \oplus l_0) X_0}\,.
\ee
The main point to notice is that the addition law $(k_0 \oplus l_0, k_a \oplus l_a)$ is clearly non-abelian if the group multiplication $g h$ is. At the algebraic level the composition law for momenta is related to the action of translation generators. Intuitively this can be seen by looking at the action of a translation generator on eigenstates in a Hilbert space i.e.\! on
states of a quantum relativistic particle of a given momentum. In this case the non-trivial composition of momentum eigenvalues can be understood as a non-Leibniz action of translaton generators on (tensor) products of states. In the language of Hopf algebras this is formalized in terms of a non-trivial {\it co-product} for space translation generators 
\be\label{copB1}
\Delta K_a = K_a \otimes 1 + e^{-K_0/\kappa} \otimes K_a\,, 
\ee
while the time translation generators exhibit a trivial co-product i.e.\! act according to the usual Leibniz rule
\be\label{copB2}
\Delta K_0 = K_0 \otimes 1 + 1 \otimes K_0\,.
\ee
Analogously the group inversion is reflected in a non-trivial {\it antipode} for the generators
\begin{align}\label{eq:3.1}
S(K_0) = -K_0\,, \qquad S(K_a) = -e^{K_0/\kappa} K_a\,,
\end{align}
which determines the bicrossproduct coordinates of the inverse group element $g^{-1}$. The relations above can be used to infer the co-products and antipodes for translation generators associated with cartesian coordinates on $AN(n)$. These generators are known in the literature as the ``classical" basis \cite{Ruegg:1994}; in the Lorentzian case we have
\begin{align}\label{eq:3.2}
\Delta P_0 & = P_0 \otimes \frac{1}{\kappa} (P_0 + P_{-1}) + \kappa (P_0 + P_{-1})^{-1} \otimes P_0 + (P_0 + P_{-1})^{-1} P_a \otimes P^a\,, \nonumber\\ 
\Delta P_a & = P_a \otimes \frac{1}{\kappa} (P_0 + P_{-1}) + 1 \otimes P_a\,, \nonumber\\ 
S(P_0) & = -P_0 + (P_0 + P_{-1})^{-1} P_aP^a\,, \qquad S(P_a) = -\kappa (P_0 + P_{-1})^{-1} P_a\,,
\end{align}
with $P_{-1} = (\kappa^2 + P_0^2 - P_aP^a)^{\frac{1}{2}}$ (see also the derivations in \cite{Nowak:2002} and \cite{Govindarajan:2009,Kovacevic:2012,Borowiec:2014,Borowiec:2013}) while in the Euclidean case we can write
\begin{align}\label{eq:3.3}
\Delta P_0 & = P_0 \otimes \frac{1}{\kappa} (P_0 + P_{-1}) + \kappa (P_0 + P_{-1})^{-1} \otimes P_0 - (P_0 + P_{-1})^{-1} P_a \otimes P^a\,, \nonumber\\ 
\Delta P_a & = P_a \otimes \frac{1}{\kappa} (P_0 + P_{-1}) + 1 \otimes P_a\,, \nonumber\\ 
S(P_0) & = -P_0 - (P_0 + P_{-1})^{-1} P_aP^a\,, \qquad S(P_a) = -\kappa (P_0 + P_{-1})^{-1} P_a\,,
\end{align}
with $P_{-1} = (\kappa^2 + P_0^2 + P_aP^a)^{\frac{1}{2}}$. What we described so far is (part of) the non-trivial {\it co-algebra} structure of two bases of translation generators associated with the $AN(n)$ group manifold. Notice that when the group momentum space ``flattens" to ordinary Minkowski space i.e.\! when $\kappa \rightarrow +\infty$ we recover the ordinary action of Poincar\'e generators.\\
A question emerges at this point: how are the Lie {\it algebra} structure of translations and Lorentz generators and the {\it co-algebra} structure of Lorentz generators affected by the ``deformations" described above? It turns out that (\ref{copB1}), (\ref{copB2}) and (\ref{eq:3.2}) describe the translation sector, in two different ``bases", of the $\kappa$-Poincar\'e Hopf algebra, a {\it quantum deformation} of the Poincar\'e algebra introduced in \cite{Lukierski:1991} (see also \cite{Lukierski:1994}). The co-products and antipodes in (\ref{eq:3.3}) should be seen as the Euclidean counterpart of such translation sector. The generators of the $\mathfrak{an}(n)$ algebra $X_0$ and $X_a$ are interpreted as time and spatial positions, respectively, spanning the $\kappa$-Minkowski non-commutative spacetime \cite{Majid:1994}. In order to answer the questions above let us focus on the 3+1-dimensional case. For the generators $P_0$, $P_a$ the Lie algebra structure can be determined from the Lorentz transformations of the embedding space of the momentum manifold \cite{Kowalski:2002}. The result shows that the commutators of Lorentz and  translation generators reproduce the ones of the ordinary Poincar\'e algebra, from which the denomination ``classical" basis. Transforming to the bicrossproduct basis it can be shown that the commutator between translation and boost generators is replaced by the following deformed relation  
\begin{align}\label{eq:3.5}
[K_0, N_a] & = -i N_a\,, \nonumber\\ 
[K_a, N_b] & = -i\delta_{ab} \left(\frac{\kappa}{2} \left(1 - e^{-2 K_0/\kappa}\right) + \frac{1}{2\kappa} K_a K^a\right) + \frac{i}{\kappa} K_a K_b
\end{align}
while all other commutators remain unmodified. The non-linearities appearing in the $\kappa$-deformed commutator lead to finite boost transformations in which $\kappa$ plays the role of an invariant fundamental energy scale providing an example of the so-called ``doubly special relativity" kinematics \cite{AmelinoCamelia:2000mn,Kowalski:2002,Kowalski:2003}.\\
At the co-algebra level the $\kappa$-Poincar\'e algebra is characterized by ordinary (i.e.\! trivial co-products and antipodes) generators of rotations while boost generators are deformed, to wit
\begin{align}\label{eq:3.6}
\Delta M_a & = M_a \otimes 1 + 1 \otimes M_a\,, \qquad S(M_a) = -M_a\,, \nonumber\\ 
\Delta N_a & = N_a \otimes 1 + e^{-K_0/\kappa} \otimes N_a + \frac{1}{\kappa} \varepsilon_{abc} K^b \otimes M^c\,, \nonumber\\ 
S(N_a) & = -e^{K_0/\kappa} N_a + \frac{1}{\kappa} \varepsilon_{abc} e^{K_0/\kappa} K^b M^c\,.
\end{align}
Notice that setting $K_0 = K_a = 0$, i.e.\! restricting to the Lorentz algebra, we recover a trivial Hopf algebra structure. The corresponding relations for the Euclidean case can be obtained \cite{Lukierski:1994} through a Wick rotation $\kappa \mapsto i \kappa$, $K_0 \mapsto i K_0$, $N_a \mapsto i N_a$ (cf.\! the previous Section), which leads to the deformed commutators
\begin{align}\label{eq:3.7}
[K_0, N_a] & = i N_a\,, \nonumber\\ 
[K_a, N_b] & = -i\delta_{ab} \left(\frac{\kappa}{2} \left(1 - e^{-2 K_0/\kappa}\right) - \frac{1}{2\kappa} K_a K^a\right) - \frac{i}{\kappa} K_a K_b\,.
\end{align}
and the co-products and antipodes 
\begin{align}\label{eq:3.8}
\Delta M_a & = M_a \otimes 1 + 1 \otimes M_a\,, \qquad S(M_a) = -M_a\,, \nonumber\\ 
\Delta N_a & = N_a \otimes 1 + e^{-K_0/\kappa} \otimes N_a - \frac{1}{\kappa} \varepsilon_{abc} K^b \otimes M^c\,, \nonumber\\ 
S(N_a)&  = -e^{K_0/\kappa} N_a - \frac{1}{\kappa} \varepsilon_{abc} e^{K_0/\kappa} K^b M^c\,.
\end{align}
Analogous relations for the $\kappa$-Poincar\'e algebra in the classical basis are easily obtained. Indeed for the Lorentzian signature the coproducts and antipodes of boost generators in the classical basis have the form
\begin{align}
\Delta N_a & = N_a \otimes 1 + \kappa (P_0 + P_{-1})^{-1} \otimes N_a + \varepsilon_{abc} (P_0 + P_{-1})^{-1} P^b \otimes M^c\,, \nonumber\\ 
S(N_a) & = -\frac{1}{\kappa} (P_0 + P_{-1}) N_a + \frac{1}{\kappa} \varepsilon_{abc} P^b M^c\,,
\end{align}
see also \cite{Kovacevic:2012,Borowiec:2014,Borowiec:2013}. For the Euclidean signature we obtain
\begin{align}
\Delta N_a & = N_a \otimes 1 + \kappa (P_0 + P_{-1})^{-1} \otimes N_a - \varepsilon_{abc} (P_0 + P_{-1})^{-1} P^b \otimes M^c\,, \nonumber\\ 
S(N_a) & = -\frac{1}{\kappa} (P_0 + P_{-1}) N_a - \frac{1}{\kappa} \varepsilon_{abc} P^b M^c\,.
\end{align}
This completes our short review of the structure of the Lorentzian $\kappa$-Poincar\'e algebra in the classical and bicrossproduct basis. We also derived in this Section the relations for the Euclidean $\kappa$-Poincar\'e algebra in the classical basis for the first time. Below we will study how these deformations of the algebraic structures of ordinary relativistic kinematics reflect on the behaviour of the spectral dimension providing some insight on the effective geometry associated with them.

\section{The spectral dimension of $\kappa$-Minkowski space}
The spectral dimension is a measure of the effective dimensionality of a given Riemannian manifold $(M,h)$ ($h$ denotes the metric), which is introduced in the following way. We consider a ficticious diffusion process on the manifold, governed by the heat equation
\begin{align}\label{eq:4.1}
\frac{\partial}{\partial \sigma} K(x,x_0; \sigma) = \Delta_h K(x,x_0; \sigma)\,, \quad K(x,x_0; 0) = \frac{\delta(x - x_0)}{\sqrt{\det g(x)}}\,,
\end{align}
where $\sigma$ is a ``diffusion time" parameter and the Laplacian $\Delta_h$ depends on the metric. Its solution can be characterized by the average return probability
\begin{align}\label{eq:4.2}
{\cal P}(\sigma) = \frac{1}{V} \int\! d^dx\ \sqrt{\det h(x)} K(x,x; \sigma)\,,
\end{align}
where $V$ denotes the volume, which in principle may go to infinity. The spectral dimension can be extracted by taking the logarithmic derivative of the return probability
\begin{align}\label{eq:4.3}
d_S = -2 \frac{d\log {\cal P}(\sigma)}{d\log\sigma}\,.
\end{align}
In order to apply this concept to a pseudo-Riemannian manifold one first has to find its Euclidean counterpart. A solution of the heat equation can also be given in the momentum representation, in the form
\begin{align}\label{eq:4.4}
K(x,x_0; \sigma) = \int\! d^dp\ e^{-\sigma C(p)} e^{i p (x - x_0)}\,,
\end{align}
where $C(p)$ is a momentum space version of the Laplacian. In particular, one may use this framework for the $\kappa$-Minkowski space-time \cite{Benedetti:2009}. For definiteness we start by giving a formula for the return probability in the classical basis of the $\kappa$-Poincar\'e algebra in 3+1 dimensions, based on our results from the Section II,
\begin{align}\label{eq:4.5}
{\cal P}(\sigma) = \int\! d^5p\ \delta\left(p_{-1}^2 - (p_0^2 + p_ap^a + \kappa^2)\right) \theta(p_{-1} - \kappa) e^{-\sigma C(p_0, \{p_a\})} = \nonumber\\ 
\int d^4p \frac{1}{2 \sqrt{p_0^2 + p_ap^a + \kappa^2}} e^{-\sigma C(p_0, \{p_a\})}\,,
\end{align}
where the integral is taken over the five-dimensional Minkowski embedding space and $\theta(.)$ denotes the Heaviside function. Various choices of the Laplacian function $C(p_0, \{p_a\})$, the $\kappa$-deformed ``mass-shell", are possible which are dictated by different requirements. Here we will explore three possible Laplacians which have appeared or are suggested by approaches to $\kappa$-deformations in the literature and calculate the associated spectral dimension. We perform the calculation for the $AN(2)$ and $AN(3)$ momentum group manifolds, which are associated with three and four dimensional $\kappa$-deformations of the Poincar\'e group, respectively.

\subsection{Bi-covariant Laplacian}
We start with the Laplacian obtained from the Euclidean version of the $\kappa$-deformed mass Casimir 
\begin{align}\label{eq:4.6}
C_1(K_0,\{K_a\}) = 4\kappa^2 \sinh^2\left(\tfrac{K_0}{2\kappa}\right) + 
e^{K_0/\kappa} K_a K^a + \nonumber\\ 
\frac{1}{4\kappa^2} \left(4\kappa^2 \sinh^2\left(\tfrac{K_0}{2\kappa}\right) + 
e^{K_0/\kappa} K_a K^a\right)^2\,.
\end{align}
This Casimir can be seen as the momentum space counterpart of the Laplacian associated with the so-called {\it bi-covariant differential calculus} \cite{Ruegg:1994}. In terms of cartesian coordinates it leads to an undeformed mass shell i.e.
\be\label{eq:4.6a}
C_1(p_0, \{p_a\}) = p_0^2 + p_ap^a = p_{-1}^2 - \kappa^2\, .
\ee
The return probability is then given by the integral
\begin{align}\label{eq:4.7}
{\cal P}(\sigma) = \int\! d^4p\ \frac{1}{2 \sqrt{p_0^2 + p_ap^a + \kappa^2}}\, 
e^{-\sigma (p_0^2 + p_ap^a)}\,.
\end{align}
From this one calculates that in 3+1 dimensions i.e.\! for $AN(3)$ the spectral dimension is given by the expression
\begin{align}\label{eq:4.8}
d_S = \frac{2\kappa \sqrt{\sigma} (2\kappa^2 \sigma - 3) - 
\sqrt{\pi} e^{\kappa^2 \sigma} (4\kappa^4 \sigma^2 - 4\kappa^2 \sigma + 3) \mathrm{erfc}(\kappa \sqrt{\sigma})}{-2\kappa \sqrt{\sigma} + \sqrt{\pi} e^{\kappa^2 \sigma} (2\kappa^2 \sigma - 1) \mathrm{erfc}(\kappa \sqrt{\sigma})}\,,
\end{align}
where $\mathrm{erfc}(.)$ is the so-called ``complementary error function". We can take the UV and IR limits of the expression above, related to diffusion at small $(\sigma \ll 1/\kappa^2)$ and large scales $(\sigma \gg 1/\kappa^2)$, which are given by
\begin{align}\label{eq:4.9}
\lim_{\sigma \to 0} d_S = 3\,, \quad \lim_{\sigma \to +\infty} d_S = 4\,, 
\end{align}
which reproduces the result obtained numerically in \cite{Benedetti:2009}. In turn, in 2+1 dimensions i.e.\! for $AN(2)$ the spectral dimension is given by
\begin{align}\label{eq:4.10}
d_S = 2 + \frac{\kappa^2 \sigma U(\frac{3}{2},1,\kappa^2 \sigma)}
{U(\frac{1}{2},0,\kappa^2 \sigma)}\,,
\end{align}
where $U(.,.,.)$ is a Tricomi confluent hypergeometric function. Again we can take the UV and IR limits
\begin{align}\label{eq:4.11}
\lim_{\sigma \to 0} d_S = 2\,, \quad \lim_{\sigma \to +\infty} d_S = 3\,, 
\end{align}
showing a pattern similar to the four-dimensional case (as suggested in \cite{Benedetti:2009ge}). The results in both cases are plotted in Fig.\! 3. Notice that the flow of the spectral dimension for the Laplacian $C_1$ is exclusively due to the non-trivial integration measure on the momentum group manifold in the expression for the return probability. In the cases explored below we will see that further contribution will be given by the presence of a deformed Laplacian.
\begin{figure}[ht]\label{fig:dsc1}
\includegraphics[width=0.4\textwidth]{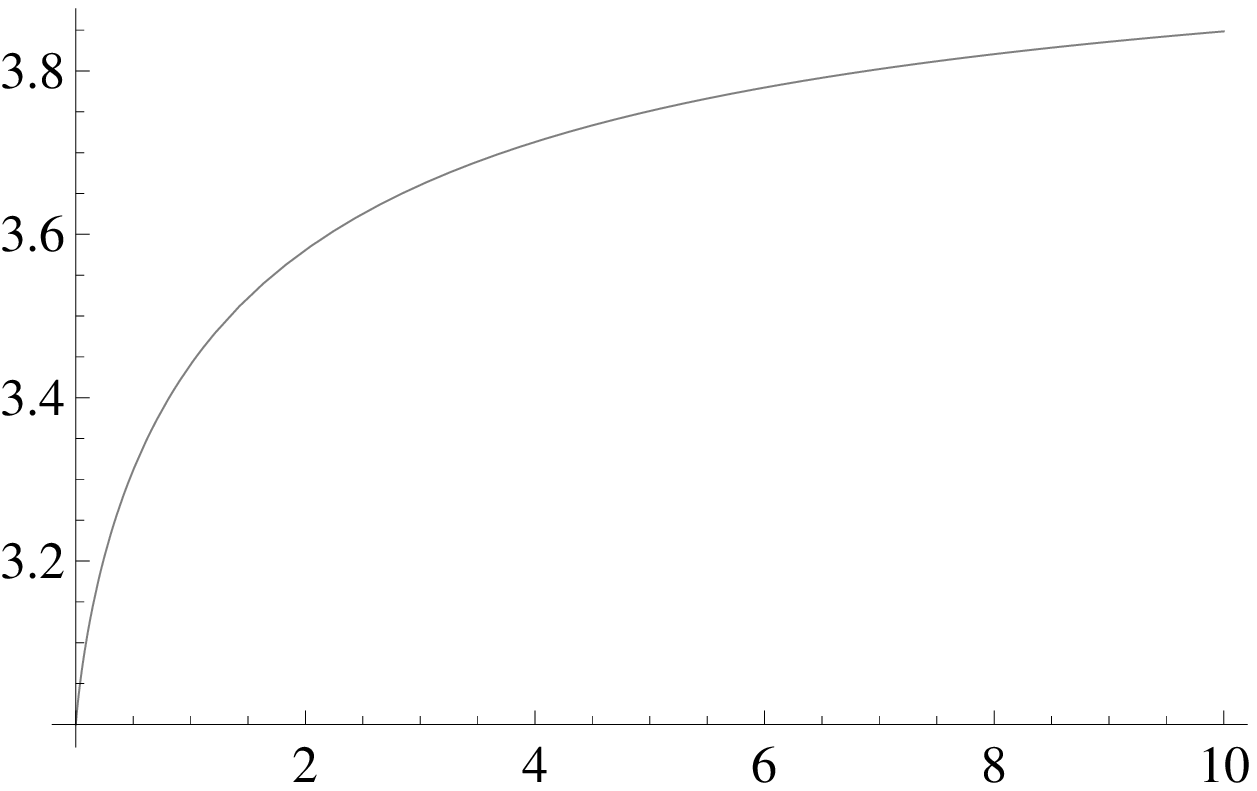}\qquad\qquad\,\,
\includegraphics[width=0.4\textwidth]{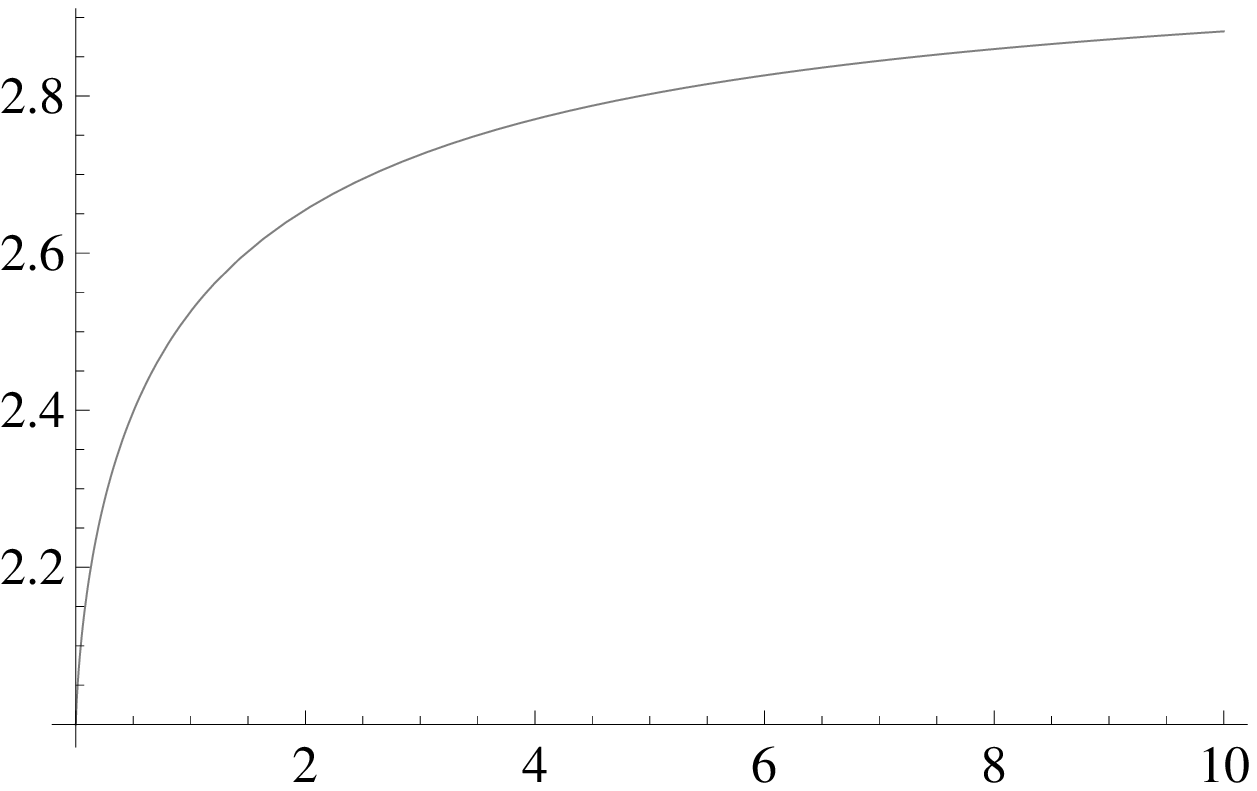}
\caption{Spectral dimension $d_S(\sigma)$ for $C_1$ in 3+1 dim (left) and in 2+1 dim (right) for $\kappa = 1$}
\end{figure}

\subsection{Bi-crossproduct Laplacian}
We consider now the Euclidean version of the natural mass Casimir for the $\kappa$-Poincar\'e algebra in the bi-crossproduct basis \cite{Lukierski:1992} 
\begin{align}\label{eq:4.12}
C_0(K_0,\{K_a\}) = 4\kappa^2 \sinh^2\left(\tfrac{K_0}{2\kappa}\right) + e^{K_0/\kappa} K_aK^a\,.
\end{align}
The mass-shell determined by such invariant in cartesian coordinates is given by
\be\label{eq:4.12a}
C_0(p_0, \{p_a\}) = 2\kappa \left(\sqrt{p_0^2 + p_ap^a + \kappa^2} - \kappa\right) = 2\kappa (p_{-1} - \kappa)\,.
\ee
The integral defining the return probability is now given by
\begin{align}\label{eq:4.13}
{\cal P}(\sigma) = \int\! d^4p\ \frac{1}{2 \sqrt{p_0^2 + p_ap^a + \kappa^2}}\, 
e^{-2\kappa \sigma (\sqrt{p_0^2 + p_ap^a + \kappa^2} - \kappa)}\,.
\end{align}
We can analytically evaluate this integral in 3+1 dimensions to obtain the following simple expression for the spectral dimension
\begin{align}\label{eq:4.14}
d_S = \frac{8\kappa^2 \sigma + 6}{2\kappa^2 \sigma + 1}\,.
\end{align}
The asymptotic values in the IR and UV are easily calculated to be
\begin{align}\label{eq:4.15}
\lim_{\sigma \to 0} d_S = 6\,, \quad \lim_{\sigma \to +\infty} d_S = 4\,, 
\end{align}
reflecting the asymptotic properties already found in the recent literature \cite{Amelino:2013}. In 2+1 dimensions the analytical evaluation of the return probability does not seem possible and we resorted to numerical methods. The resulting plot for the spectral dimension is given in Fig.\! 4 and its asymptotic values in the UV and IR are found to be
\begin{figure}[ht]\label{fig:dsc0}
\includegraphics[width=0.4\textwidth]{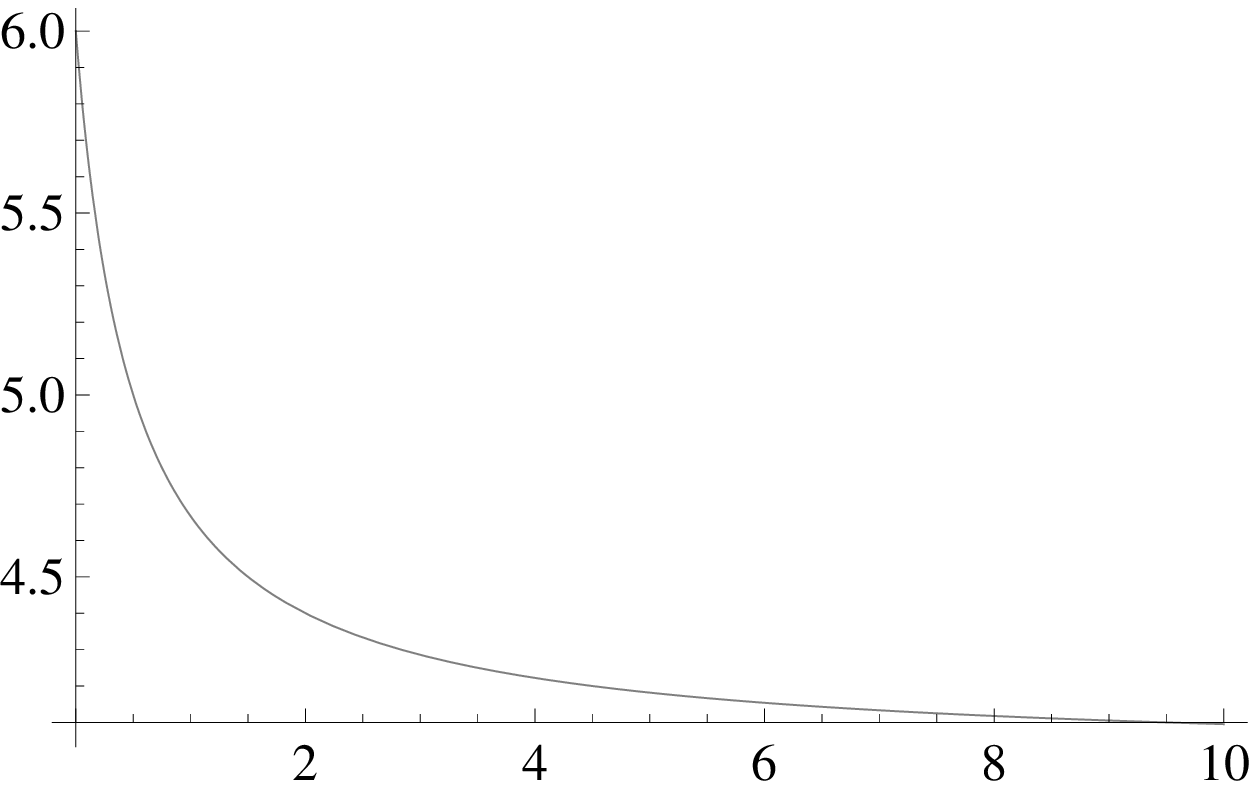}\qquad\qquad\,\,
\includegraphics[width=0.4\textwidth]{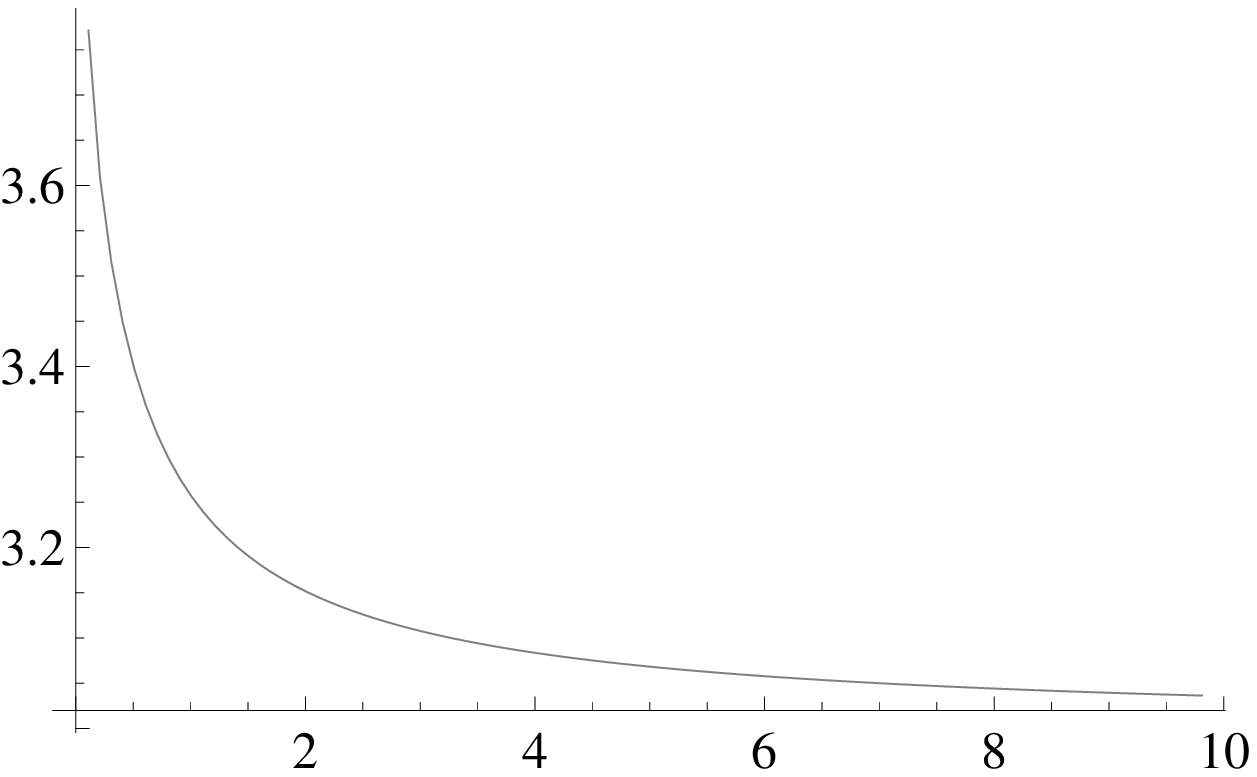}
\caption{Spectral dimension $d_S(\sigma)$ for $C_0$ in 3+1 dim (left) and in 2+1 dim (right) for $\kappa = 1$}
\end{figure} 
\begin{align}\label{eq:4.16}
\lim_{\sigma \to 0} d_S \approx 4\,, \quad \lim_{\sigma \to +\infty} d_S \approx 3\,, 
\end{align}
showing again a similar pattern to the higher dimensional counterpart. \\

\subsection{``Relative Locality" Laplacian} 
The $AN(n)$ Lie group momentum space characterizing the $\kappa$-Poincar\'e algebra falls under the class of theories contemplated in the recent approach to deformed Planck-scale kinematics based on curved momentum space and dubbed as ``relative locality" \cite{Amelino:2011}. This approach puts strong emphasis on the geometric properties of the momentum manifold and in particular suggests that the kinematical mass Casimir invariant is determined by the (square of the) geodesic distance from the origin in the momentum space. The distance along timelike geodesics in de Sitter spacetime and hence in the Lorentzian $\kappa$-momentum space \cite{Gubitosi:2013}, which as a manifold is half of de Sitter space, is given by
\be\label{ddS}
d^2(p_0,p_a) = \kappa^2 \mathrm{arccosh}^2 (p_{-1}/\kappa) = -\kappa^2 \mathrm{arccos}^2 (p_{-1}/\kappa)\,.
\ee
Naively, one could follow the path leading from the Lorentzian to Euclidean space of momenta (\ref{eq:2.6}) and make the Wick rotation $\kappa \mapsto i \kappa$, $p_{-1} \mapsto i p_{-1}$, which yields $d^2(p_0,p_a) = \kappa^2 \mathrm{arccos}^2 (p_{-1}/\kappa)$. However, this is in fact the distance along timelike geodesics in anti-de Sitter spacetime, while our Euclidean momentum space is the hyperbolic space, which is a spatial slice of a one dimension higher anti-de Sitter space. Therefore, we should instead consider the distance along spacelike geodesics in anti-de Sitter $d^2(p_0,p_a) = \kappa^2 \mathrm{arccosh}^2 (p_{-1}/\kappa)$, which has actually the same form as the distance (squared) in (\ref{ddS}) above. In bicrossproduct coordinates the deformed mass-shell obtained from the distance (\ref{ddS}) is thus
\begin{align}\label{eq:4.17}
C_d(k_0,\{k_a\}) = \kappa^2 \mathrm{arccosh}^2 \left(\cosh\left(\tfrac{k_0}{\kappa}\right) + \frac{1}{2\kappa^2} e^{k_0/\kappa} k_ak^a\right)\,.
\end{align}
The return probability integral in cartesian coordinates is given by
\begin{align}\label{eq:4.18}
{\cal P}(\sigma) = \int\! d^4p\ \frac{1}{2 \sqrt{p_0^2 + p_ap^a + \kappa^2}}\, 
e^{-\kappa^2 \sigma \mathrm{arccosh}^2 (1/\kappa \sqrt{p_0^2 + p_ap^a + \kappa^2})}\,,
\end{align}
where we used the relation $d^2(p_0,p_a) = \kappa^2\, \mathrm{arccosh}^2 (1/\kappa \sqrt{p_0^2 + p_ap^a + \kappa^2})$. In order to compute the integral in both 3+1 and 2+1 dimensions one has to resort to numerical techniques. The results are plotted in Fig.\! 5. 
\begin{figure}[ht]\label{fig:dscd}
\includegraphics[width=0.4\textwidth]{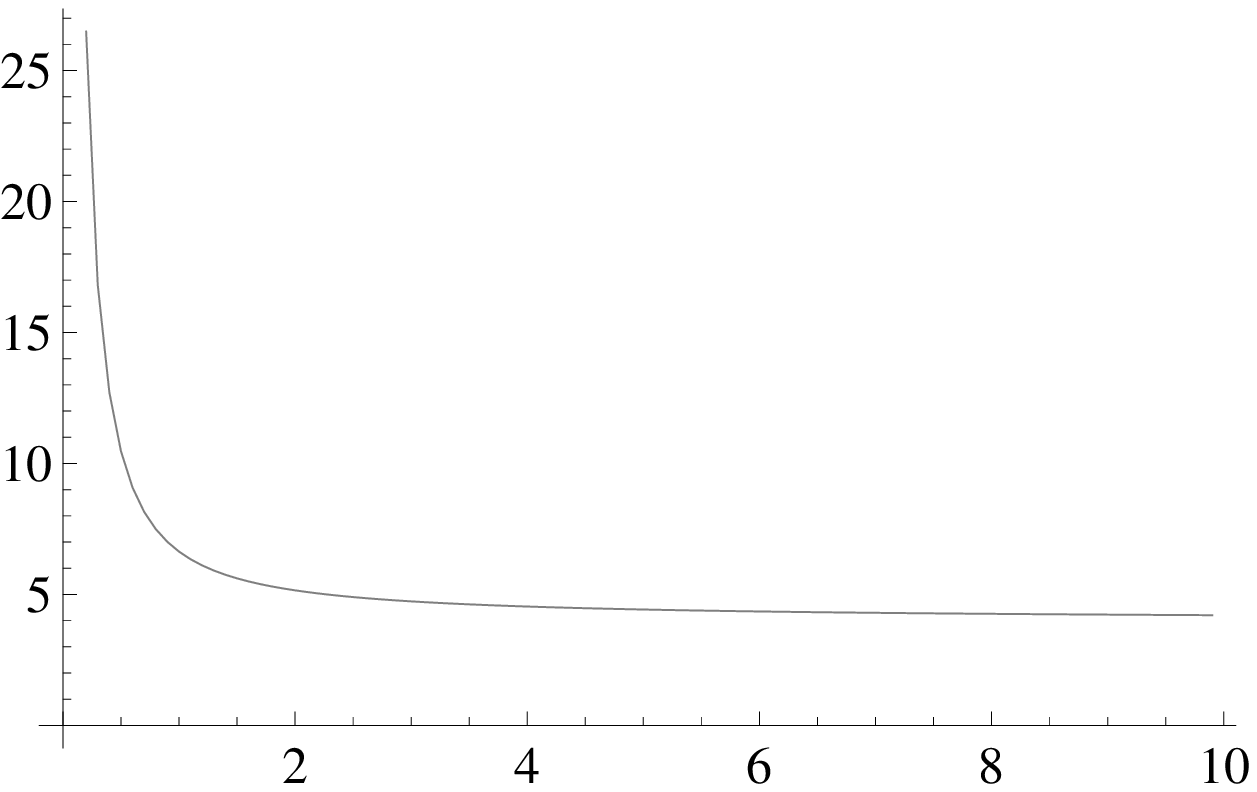}\qquad\qquad\,\,
\includegraphics[width=0.4\textwidth]{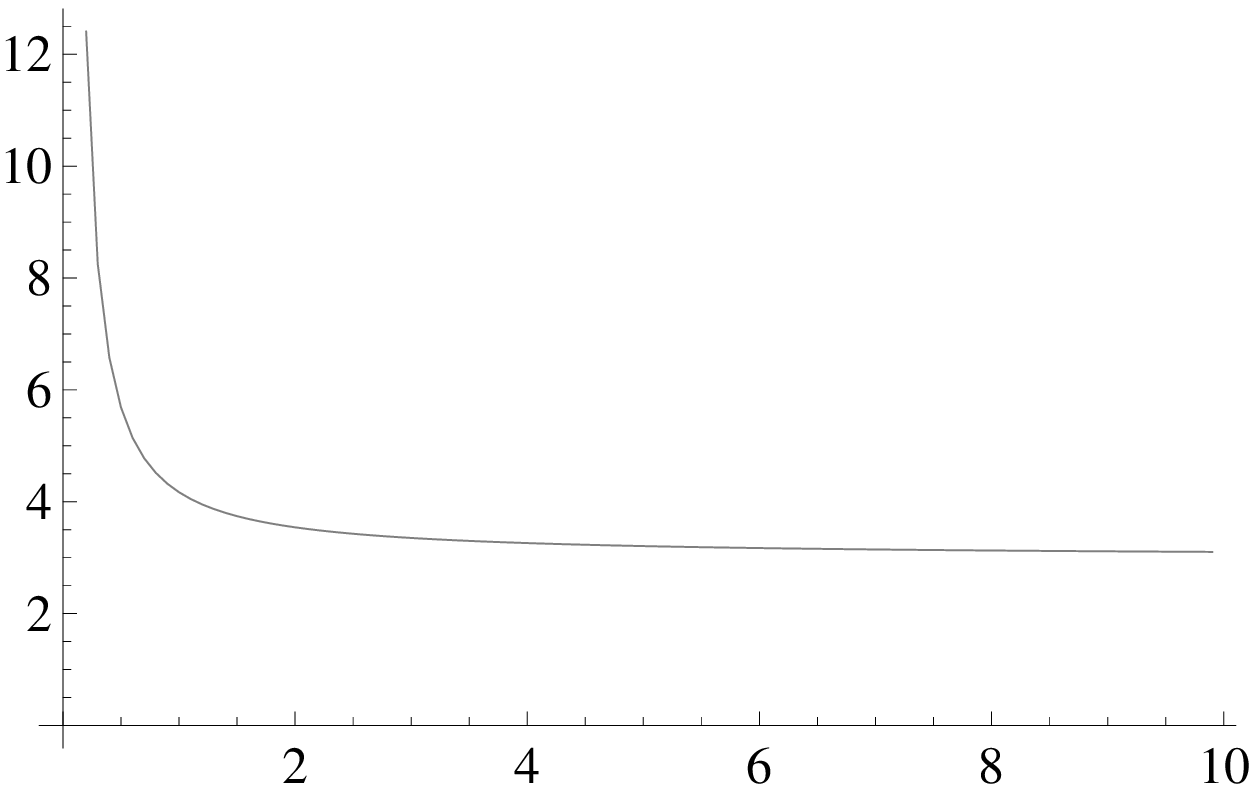}
\caption{Spectral dimension $d_S(\sigma)$ for $C_d$ in 3+1 dim (left) and in 2+1 dim (right) for $\kappa = 1$}
\end{figure}
The asymptotic values of the spectral dimension can only be computed numerically and in the four-dimensional case are given by
\begin{align}\label{eq:4.19}
\lim_{\sigma \to 0} d_S \approx +\infty\,, \quad \lim_{\sigma \to +\infty} d_S \approx 4\,, 
\end{align}
while in the three dimensional case
\begin{align}\label{eq:4.19a}
\lim_{\sigma \to 0} d_S \approx +\infty\,, \quad \lim_{\sigma \to +\infty} d_S \approx 3\,, 
\end{align}
thus in both cases the spectral dimension {\it diverges} to infinity as the diffusion time approaches zero. Below we summarize our results.

\section{Discussion}
The aim of the present work was to present a thorough study of the spectral dimension of $\kappa$-Minkowski when the associated group manifold momentum space is three- or four-dimensional. As a first step we introduced a way of defining Euclidean $\kappa$-momentum space and the corresponding generators of the $\kappa$-Poincar\'e algebra based on geometric construction which does not rely on a specific choice of basis of generators and/or momentum coordinates. We then studied the return probability and spectral dimension associated with diffusion processes based on three possible forms of the Laplacian proposed in the literature. These Laplacians appear in the integral defining the return probability in the form of mass-shell constraint functions on momentum. We found quite distinct behaviours for the three choices of Laplacians analyzed.\\
For the Casimir function determined by the Laplacian associated with the non-commutative bi-covariant differential calculus (Section IVa) we observed a typical pattern of {\it dimensional reduction} in which the spectral dimension monotonically decreases from the value of the Hausdorff dimension i.e.\! 4 or 3 in the IR to a smaller value in the UV. In particular for the $AN(3)$ group manifold we reproduced {\it analytically} the results derived numerically in \cite{Benedetti:2009} which relied on a algebraic Wick rotation of the algebra generators. The dimensional reduction observed could suggest, as noticed already in \cite{Benedetti:2009} a fractal nature of the space associated with a $AN(n)$ momentum space and this particular Laplacian. Furthermore, in the three-dimensional case we obtained the same value of the spectral dimension in the UV as in some other approaches to quantum gravity, namely causal dynamical triangulations, Ho\v{r}ava-Lifshitz gravity and (to some extent) spin foam models \cite{Benedetti:2009ge}.\\
For the diffusion process governed by the Laplacian represented in momentum space by the bi-crossproduct Casimir, a widely studied $\kappa$-deformed energy momentum {\it dispersion relation}, we obtained instead a spectral dimension which in the UV monotonically increases to a value higher than the Hausdorff dimension. In particular we reproduced the asymptotic result of \cite{Amelino:2013}, providing here an analytical expression for the spectral dimension in the four-dimensional case. The pattern of the spectral dimension in both three and four dimensions is that of {\it super-diffusion}, namely the non-trivial features of momentum space and Laplacian lead to a diffusion process which seems to explore more dimensions than the usual Hausdorff one.\\
Finally we provided a first analysis of the return probability and associated spectral dimension for the Casimir determined by the geodesic distance from the origin of momentum space as suggested by the framework of ``relative locality". Here we encountered an even stranger behaviour in which super-diffusion is accompanied by a {\it diverging} value of the spectral dimension in the UV. A similar divergence, albeit for a finite Planckian value of the diffusion time, has been encountered in the context of non-local Laplacians proposed in the context of effective field theories of quantum space-time reproducing the Bekenstein-Hawking black hole entropy \cite{Arzano:2013rka}. There the divergence was interpreted as a {\it breakdown} of the notion of diffusion.\\
Let us mention in closing that for the three-dimensional $AN(2)$ momentum space the behaviour of the spectral dimension for all three choices of the Laplacian is significantly different from that encountered in the case of a compact $SU(2)$ momentum group manifold analyzed in \cite{Alesci:2012} in the context of three-dimensional gravity. In particular for the same choice of Casimir/Laplacian as in Section VIa the spectral dimension in the $SU(2)$ momentum space first exhibits super-diffusion, reaching a maximum value at a given Planckian diffusion time and then drops to zero for vanishing diffusion times.\\
We hope that the results presented can contribute towards a more intuitive characterization of the properties of non-commutative spaces associated with momentum group manifolds. The picture resulting after our discussion seems to provide evidence that different choices of mass Casimir and thus energy-momentum dispersion relation can lead to dramatically different scenarios. The question of determining a criterion to single out a preferred notion of mass invariant in these type of theories thus remains of paramount importance and addressing it should be set as a priority for future work.

\section*{Acknowledgments}
We thank Prof. Kowalski-Glikman for many useful discussions. The work of MA is supported by a Marie Curie Career Integration Grant within the 7th European Community Framework Programme and in part by a grant from the John Templeton Foundation. TT acknowledges the support by the Foundation for Polish Science International PhD Projects Programme co-financed by the EU European Regional Development Fund and the additional funds provided by the National Science Center under the agreement no. DEC-2011/02/A/ST2/00294 and by the European Human Capital Program.


\begin{thebibliography}{99}

\bibitem{Ambjorn:2005db}
  J.~Ambjorn, J.~Jurkiewicz and R.~Loll,
  Phys.\ Rev.\ Lett.\ {\bf 95}, 171301 (2005)
  [hep-th/0505113].

\bibitem{Lauscher:2005qz}
  O.~Lauscher, M.~Reuter,
  JHEP {\bf 0510}, 050 (2005)
  [hep-th/0508202].

\bibitem{Horava:2009if}
  P.~Horava,
  Phys.\ Rev.\ Lett.\ {\bf 102}, 161301 (2009)
  [arXiv:0902.3657 [hep-th]].

\bibitem{Modesto:2008jz}
  L.~Modesto,
  Class.\ Quant.\ Grav.\  {\bf 26}, 242002 (2009)
  [arXiv:0812.2214 [gr-qc]].

\bibitem{Carlip:2009kf}
  S.~Carlip,
  [arXiv:0909.3329 [gr-qc]].

\bibitem{Benedetti:2009ge}
  D.~Benedetti, J.~Henson,
  Phys.\ Rev.\ {\bf D80}, 124036 (2009)
  [arXiv:0911.0401 [hep-th]].

\bibitem{Sotiriou:2011aa}
  T.~P.~Sotiriou, M.~Visser and S.~Weinfurtner,
  Phys.\ Rev.\ D {\bf 84}, 104018 (2011)
  [arXiv:1105.6098 [hep-th]].

\bibitem{Calcagni:2013vsa}
  G.~Calcagni, A.~Eichhorn and F.~Saueressig,
  Phys.\ Rev.\ D {\bf 87}, 124028 (2013)
  [arXiv:1304.7247 [hep-th]].


\bibitem{Alesci:2012}
  E.~Alesci, M.~Arzano,
  Phys.\ Lett.\ B {\bf 707}, 272 (2012)
  [arXiv:1108.1507 [gr-qc]].

\bibitem{Amelino:2013}
  G.~Amelino-Camelia, M.~Arzano, G.~Gubitosi and J.~Magueijo,
  [arXiv:1311.3135 [gr-qc]].

\bibitem{Calcagni:2010bj}
  G.~Calcagni,
  JHEP {\bf 1003}, 120 (2010)
  [arXiv:1001.0571 [hep-th]].

\bibitem{Arzano:2011yt}
  M.~Arzano, G.~Calcagni, D.~Oriti and M.~Scalisi,
  Phys.\ Rev.\ D {\bf 84}, 125002 (2011)
  [arXiv:1107.5308 [hep-th]].


\bibitem{Majid:1994}
  S.~Majid, H.~Ruegg,
  Phys.\ Lett.\ B {\bf 334}, 348 (1994)
  [hep-th/9405107].

\bibitem{Lukierski:1994}
  J.~Lukierski, H.~Ruegg and A.~Nowicki,
  J.\ Math.\ Phys.\ {\bf 35}, 2607 (1994).

\bibitem{Lukierski:1991}
  J.~Lukierski, H.~Ruegg, A.~Nowicki and V.~N.~Tolstoi,
  Phys.\ Lett.\ B {\bf 264}, 331 (1991).


\bibitem{Kowalski:2002}
  J.~Kowalski-Glikman,
  Phys.\ Lett.\ B {\bf 547}, 291 (2002)
  [hep-th/0207279].

\bibitem{Kowalski:2003}
  J.~Kowalski-Glikman, S.~Nowak,
  Class.\ Quant.\ Grav.\ {\bf 20}, 4799 (2003)
  [hep-th/0304101].

\bibitem{Nowak:2002}
  J.~Kowalski-Glikman, S.~Nowak,
  Phys.\ Lett.\ B {\bf 539}, 126 (2002)
  [hep-th/0203040].


\bibitem{Benedetti:2009}
  D.~Benedetti,
  Phys.\ Rev.\ Lett.\ {\bf 102}, 111303 (2009)
  [arXiv:0811.1396 [hep-th]].
  

\bibitem{Amelino:2011}
  G.~Amelino-Camelia, L.~Freidel, J.~Kowalski-Glikman and L.~Smolin,
  Phys.\ Rev.\ D {\bf 84}, 084010 (2011)
  [arXiv:1101.0931 [hep-th]].

\bibitem{Kowalski:2013}
  J.~Kowalski-Glikman,
  Int.\ J.\ Mod.\ Phys.\ A {\bf 28}, 1330014 (2013)
  [arXiv:1303.0195 [hep-th]].

\bibitem{Freidel:2014}
  L.~Freidel, R.G.~Leigh and D.~Minic,
  Phys.\ Lett.\ B {\bf 730}, 302 (2014)
  [arXiv:1307.7080 [hep-th]].

\bibitem{Leigh:2014}
  L.~Freidel, R.G.~Leigh and D.~Minic,
  [arXiv:1405.3949 [hep-th]].


\bibitem{KowalskiGlikman:2004tz}
  J.~Kowalski-Glikman, S.~Nowak,
  [hep-th/0411154].

\bibitem{Ruegg:1994}
  H.~Ruegg, V.~N.~Tolstoi,
  Lett.\ Math.\ Phys.\ {\bf 32}, 85 (1994)
  [hep-th/9406146].


\bibitem{Govindarajan:2009}
  T.R.~Govindarajan, K.S.~Gupta, E.~Harikumar, S.~Meljanac and D.~Meljanac,
  Phys.\ Rev.\ D {\bf 80}, 025014 (2009)
  [arXiv:0903.2355 [hep-th]].

\bibitem{Kovacevic:2012}
  D.~Kova\v{c}evi\'{c}, S.~Meljanac,
  J.\ Phys.\ A: Math.\ Theor.\ {\bf 45}, 135208 (2012)
  [arXiv:1110.0944 [math-ph]].

\bibitem{Borowiec:2014}
  A.~Borowiec, A.~Pacho\l, 
  Eur.\ Phys.\ J.\ C {\bf 74}, 2812 (2014)
  [arXiv:1311.4499 [math-ph]].

\bibitem{Borowiec:2013}
  A.~Borowiec, J.~Lukierski and A.~Pachol,
  [arXiv:1312.7807 [math-ph]].


\bibitem{AmelinoCamelia:2000mn}
  G.~Amelino-Camelia,
  Int.\ J.\ Mod.\ Phys.\ D {\bf 11}, 35 (2002)
  [gr-qc/0012051].


\bibitem{Lukierski:1992}
  J.~Lukierski, A.~Nowicki and H.~Ruegg,
  Phys.\ Lett.\ B {\bf 293}, 344 (1992).


\bibitem{Gubitosi:2013}
  G.~Gubitosi, F.~Mercati,
  Class.\ Quant.\ Grav.\ {\bf 30}, 145002 (2013)
  [arXiv:1106.5710 [gr-qc]].


\bibitem{Arzano:2013rka}
  M.~Arzano, G.~Calcagni,
  Phys.\ Rev.\ D {\bf 88}, 084017 (2013)
  [arXiv:1307.6122 [hep-th]].

\end{thebibliography}
\end{document}